\documentclass[prb,twocolumn,superscriptaddress,showpacs,floats,floatfix]{revtex4-1}
\usepackage{dcolumn,amsmath,latexsym,graphicx,amssymb}
\usepackage[usenames,dvipsnames]{color} \usepackage{multirow}
\usepackage{array}
\usepackage{xcolor}

\RequirePackage{ifpdf} % ¿latex o pdflatex?
\begin{document}

\title{Dynamic bonding influenced by the proximity of adatoms to one-atom high step edges}

\author{W. Dednam}
\affiliation{Physics Department, College of Science, Engineering and Technology, University of South Africa, Private Bag X6, Florida Park, Johannesburg 1710, South Africa}
\affiliation{Departamento de F\'\i sica Aplicada and Unidad asociada CSIC, Universidad de Alicante, Campus de San Vicente del Raspeig, E-03690 Alicante, Spain.}

\author{S. Tewari}
\affiliation{Huygens-Kamerlingh Onnes Laboratorium, Leiden University, Niels Bohrweg 2, 2333 CA Leiden, Netherlands.}

\author{E.B. Lombardi}
\affiliation{Physics Department, College of Science, Engineering and Technology, University of South Africa, Private Bag X6, Florida Park, Johannesburg 1710, South Africa}

\author{J.J. Palacios}
\affiliation{Departamento de F\'\i sica  de la Materia Condensada, Condensed Matter Physics Center (IFIMAC) and Instituto Nicol\'as Cabrera, Universidad Aut\'onoma de Madrid, Madrid E-28049, Spain}

\author{J.M. van Ruitenbeek}
\affiliation{Huygens-Kamerlingh Onnes Laboratorium, Leiden University, Niels Bohrweg 2, 2333 CA Leiden, Netherlands.}

\author{C. Sabater}
\email{Electronic mail address: carlos.sabater@ua.es}
\affiliation{Departamento de F\'\i sica Aplicada and Unidad asociada CSIC, Universidad de Alicante, Campus de San Vicente del Raspeig, E-03690 Alicante, Spain.}
\date{\today}

\begin{abstract}
Low-temperature scanning tunneling microscopy is used here to study dynamic bonding of gold atoms on surfaces under low coordination conditions.  In the experiments,  using an atomically-sharp gold tip, a gold adatom is deposited onto a gold surface with atomic precision either on the first hollow site near a step edge, or far away from it. Classical molecular dynamics simulations at 4.2 K and density functional theory calculations serve to elucidate the difference in the bonding behavior between these two different placements, while also providing information on  the crystalline classification of the STM tips based on their experimental performance.
\end{abstract}

\maketitle

\section{Introduction}
%STM
The Scanning Tunnelling Microscope (STM) \cite{Binnig82,Binnig87} is the most common scientific instrument used to obtain the topography of nanometer-sized areas  with subatomic precision \cite{Herz03,Battisti7}. When this instrument is combined with high vacuum and low temperature (LT) conditions, its capabilities  increase, also allowing for atomic manipulation \cite{ibmATOMS, Stroscio91,Crommie,Hla14,OtteManipulation}. Essentially, there are two modes of operation for the  LT-STM. Firstly, the non-contact mode, that uses the tunneling current to scan and manipulate atoms on the surface. Secondly, the contact mode where bonding between apex tip atom and adatoms on the surface is allowed. This second mode is based on a gentle atomic touch and it is used to manipulate, move, or deposit adatoms onto the surface. In both modes the integrity of the tip and the surface should be preserved  \cite{sabater2012mechanical}.
%Dynamic Boding

For atomic manipulation, the atoms on the apex of a STM tip need to establish a bond with the atoms on the surface through the sharing of electrons. The process that occurs in going from the tunneling regime to the atomic contact is called dynamic bonding, and it is usually studied through the changes in the conductance during this process. Electron transport at these atomic scales is usually described by Landauer formalism \cite{Landauer} where the quantum of conductance, $G_0=2e^2/h$ ($e$ is the charge of the electron and $h$ is Planck's constant) acts as a the reference quantity.  Depending on the nature of the electrodes, the conductance can exhibit a discontinuity from the last point of tunneling to the contact point, a phenomenon which is named “jump-to-contact” (JC). Other than the technique of landing the STM tip over  target adatoms or molecules \cite{Limot2005,Krog2008Rev}, this phenomenon has also been extensively reported in experiments using the mechanically controlled break junction (MCBJ) technique \cite{mcbj92,Krans93,Atindra22} and the STM in the Break-Junction approach \cite{Agrait1993,Pascual1993,Agrait2003,Trouwborst2008,PlasticityPb,J2CUntiedt,Sabater2013,Rey18,Sabater2018}.

In this work we  use the LT-STM technique to manipulate adatoms at will so that we can study the JC phenomenon in controlled environments. In particular, we have considered two scenarios where the STM tip is used to probe an adatom (1) located in a hollow site on a flat Au(111) surface  and (2) placed close to a one-atom high step edge. We complement the LT-STM-BJ experimental results with classical molecular dynamics (CMD) simulations and density functional theory calculations for quantum transport. Although related theoretical results have been previously reported \cite{Fernandez2016}, the triple combination of these techniques is  used here for the first time to get a full picture of  electronic and mechanical properties of adatoms close and far away from the edge-steps. 
%Moreover, independently of the scenario, we have introduced the novelty to simulate via CMD continuous cycles of rupture and formation of the bonding.  
Further, our combination of techniques allows to learn about the integrity of the tip in the experiments,  offering a non-destructive new approach  to get information about the tips' geometry which is \textit{a priori} impossible to obtain just from the experiments. Understanding the tip deformation allows new insights in the field of atomic manipulation, and molecular electronics. On the other hand,  the detailed study of the deformation of the surface during the dynamic bonding process opens
the door to unmasking mechanisms and understanding effects in the emerging field of straintronics.

\section{Experimental Approach}
An ultra high-vacuum low-temperature STM in the break junction configuration is used in this work to control the motion and manipulate the position of a single adatom on a surface and measure the electronic transport. To improve the control of the STM tip and the ad-atom, a real-time molecular dynamics simulator and a 3D motion control system have been incorporated in this experimental setup \cite{TewariMicro,Tewari2018,tewari2019intuitive}. To emulate the experiment and permit synchronization with the 3D motion system, there are only a few atoms interacting via low computational cost potentials in the real-time molecular dynamics simulator.

The surface sample consists of mono-crystalline gold cut along the (111) crystallographic direction. It is prepared together with the Au tip
by repeated argon sputtering and thermal annealing cycles, in order to obtain an atomically flat Au(111) facet showing herringbone surface reconstruction. We further prepare the surface at low temperature through the creation of a localized stress pattern \cite{Oliver2014, Moresco2003, Moresco2004, Moresco2005} using gentle indentation of the STM tip at a spot on the surface removed from the area of investigation. This controlled crash of the tip onto the surface produces new crystalline (111) facets, and straight step edges in the three crystallographic directions of Au(111). Furthermore,  to make an atomically sharpened gold tip we follow the procedure detailed in Refs.~\onlinecite{sabater2012mechanical,castellanostip2012,Sumit2017b}. Additionally, gold atoms are deposited \cite{Saw2014,Yang2016,Sumit2017b,Tewari2018} on the Au(111) surface at the target sites of investigation (Figure \ref{fig2:sceneraio_STM}(c)), and the procedure by which they are manipulated is described briefly below. 

To move adatoms near these one-atom high step edges, we need to know the configuration of the substrate atoms close to the edge. The atomic configuration of metallic surfaces like Au(111) is not readily available in STM images due to the delocalized nature of the valence electrons in metals \cite{Wintterlin1989}. However, following the protocol designed by Tewari \textit{et al}.  \cite{Tewari2018,Sumit2017b} and appealing to geometric arguments, it is possible determine the orientation of the Au(111) substrate.
Once the substrate atomic configuration is known, an adatom can be moved to the step edge using the point-contact pushing (PCP) technique,  which involves maintaining contact between the adatom and tip, while the tip is lowered from the point of first contact by approximately a quarter of the atomic height, with the help of the 3D motion control system.

\begin{figure}[ht!]
\includegraphics[width=0.5\textwidth]{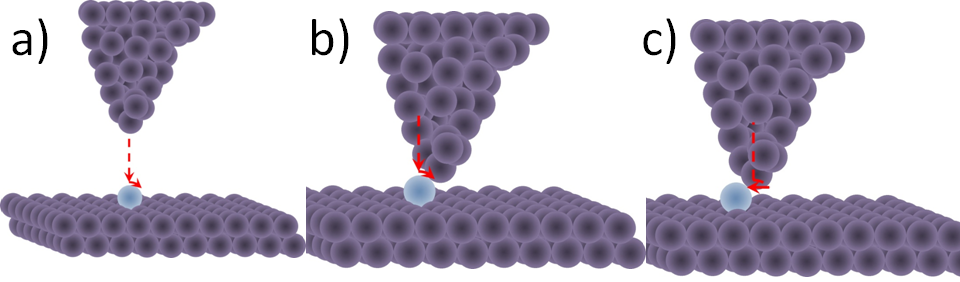}
\caption{Point contact push (PCP) technique: (a) The STM tip is positioned directly above the adatom before a point contact is made and then brought vertically downwards. (b) Once contact has been made the STM tip is moved in a circular trajectory while remaining in contact with the adatom until the tip is behind it. (c) When the tip is positioned behind the adatom it pushes the latter to the next hollow site.}
\label{fig1:PC}
\end{figure}

\begin{figure}[ht!]
\centering
\includegraphics[width=0.5\textwidth]{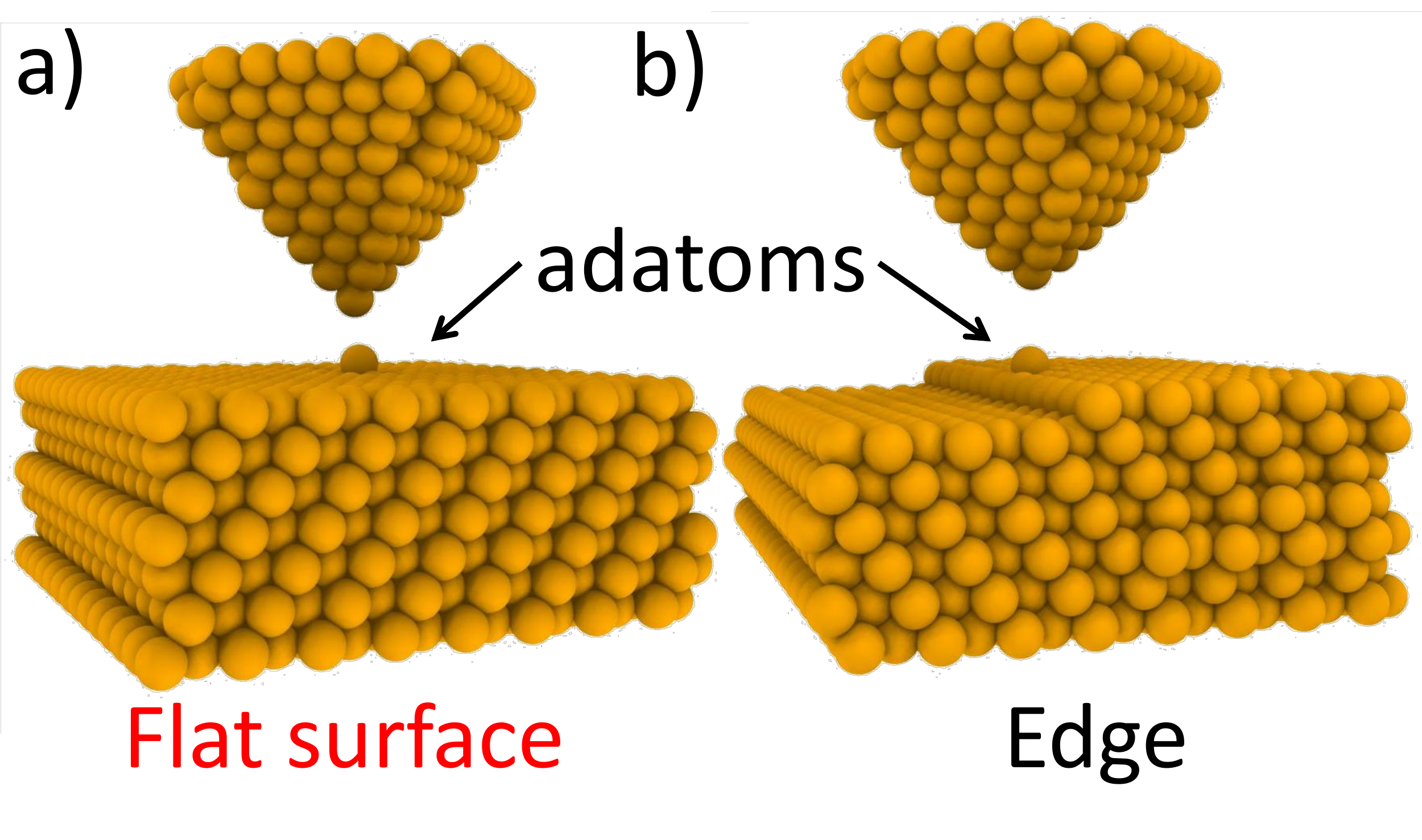}
\includegraphics[width=0.5\textwidth]{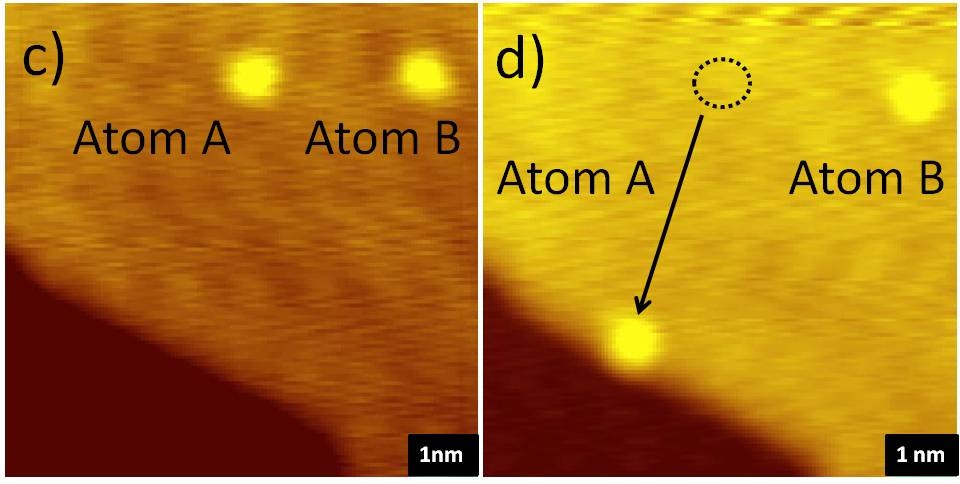}
\caption{Illustration of a STM tip and an adatom on a surface (a), all composed of gold atoms, and (b), the same STM tip and adatom close to a step edge that is one atom high. (c) STM image of the surface with 2 adatoms labeled  A and B. (d) STM image after the new position of the A atom which was moved in the first hollow directly above the step edge.} 
\label{fig2:sceneraio_STM}
\end{figure}

As shown in Figure~\ref{fig1:PC}, the tip starts over the adatom (a), and is lowered into the pushing position (b), and the adatom is pushed from one hollow site to another (c), while obtaining visual feedback from the real-time simulation, which helps to precisely place the adatoms at a target site near the step edge.

The dynamic bonding can be studied via STM-BJ experiments\cite {Agrait1993,Pascual1993,Agrait2003},  for adatom positions in the middle of the surface, or just in the first hollow site after  the step edge. In the Introduction we defined dynamical bonding as the process that occurs in going from the tunneling to the atomic contact regimes.  This dynamical bonding is observed in the form of a `jump-to-contact' (JC) \cite{J2CUntiedt}.
The JC is manifested, notably for noble metals,  when atomically sharp electrodes approach each other, and the measured conductance jumps from the tunneling regime ($ \lesssim 10^{-1}G_{0}$) to atomic contact ($ \approx G_{0}$). 

In the STM-BJ experiments carried out in this work,  after completing the PCP maneuver, placing the adatoms at their respective sites, the tip is then located directly above the adatom and slowly lowered towards it. During this process the current that flows as the tip approaches the surface is measured. In the experiments this current was amplified 
 and converted to voltage by a Femto I/V (model DLPCA-200) using the amplification $10^6$\,V/A. Knowing that the bias voltage applied over the tip and surface is 100 mV, the conductance can be expressed as $G=1/R=V/I$, and converted to quantum of conductance units, with  $G_{0}=\frac{1}{12906} \Omega^{-1}$. The curve of conductance versus the voltage applied to the piezo system is called a contact formation or contact breaking trace. In this study of the JC we focus on contact formation traces. 

\section{Experimental results}

To reproduce the scenarios illustrated in Figs. \ref{fig2:sceneraio_STM} a) and b), we have evaporated gold adatoms onto a surface of gold oriented in the (111) crystallographic direction with step edges as shown in Fig. \ref{fig2:sceneraio_STM} c). 

The scan in Fig. \ref{fig2:sceneraio_STM}(c) shows that there are two adatoms (atoms A and B) close to each other on an Au(111) facet. Through use of the PCP technique, we moved atom A from its original position to the new position at the first hollow site directly above the step edge, as shown in Fig. \ref{fig2:sceneraio_STM}(d), where the black circle indicates the original position of atom A, and an arrow its trajectory from the old to the new position close to the edge. To refer to the states of adatoms A and B in Fig. \ref{fig2:sceneraio_STM}(d), we have used the nomenclature of ``Edge'' for A and adatom on the ``Flat surface'' for B.

Once the atoms are positioned at the edge  or in the Flat surface region far from the Edge, we proceed to study the dynamic bonding of  each of these adatoms to the tip by means of the JC phenomenon.  We recorded traces of contact formation starting from the tunneling regime until contact was established between a single atom of the tip and the respective adatom (which occurs at $\approx 1 G_{0}$).

\begin{figure}[ht!]
\centering
\includegraphics[width=0.5\textwidth]{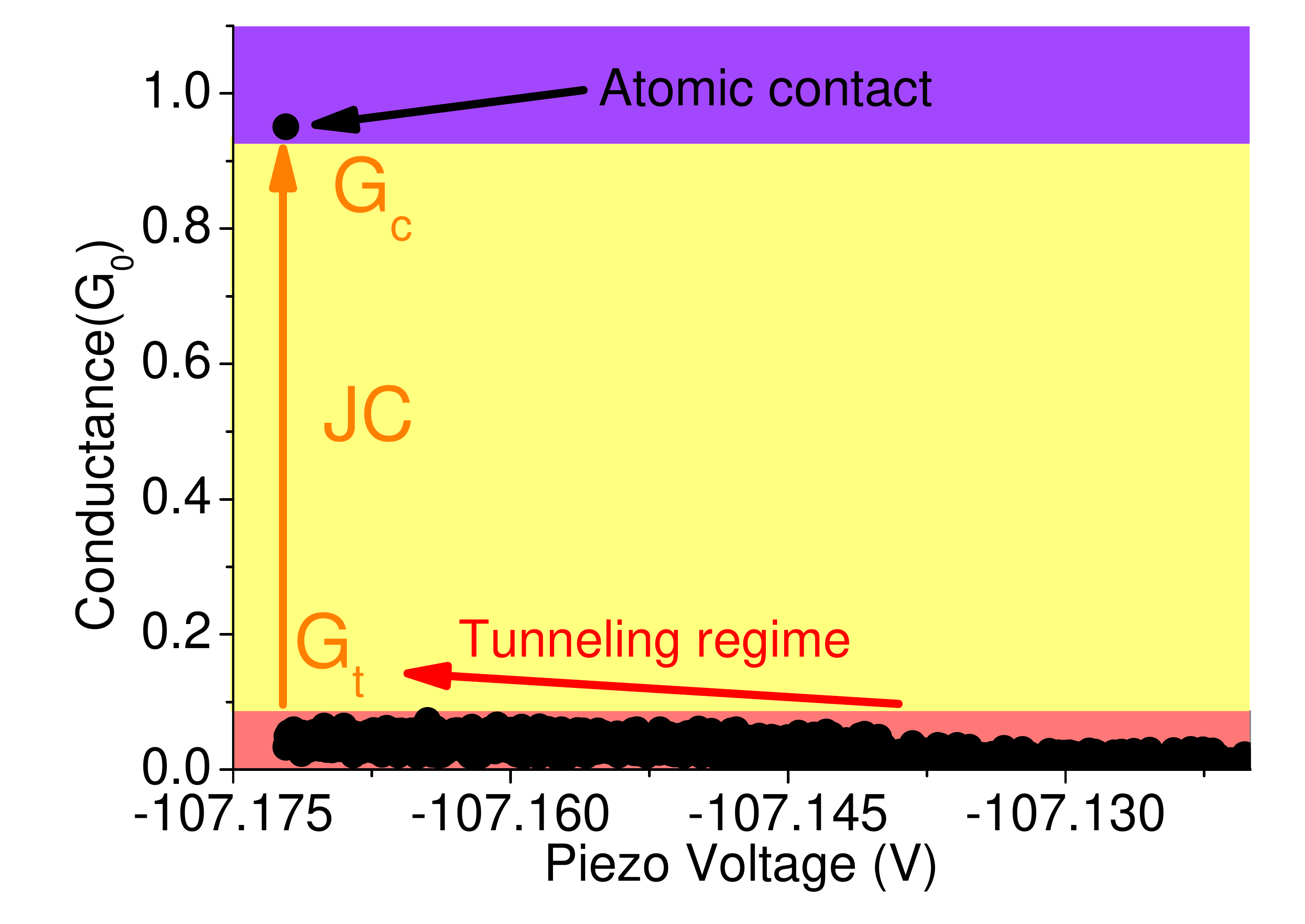}
\caption{Experimental trace of conductance for adatom on a Edge.} 
\label{fig3:exp_trace}
\end{figure}

Figure \ref{fig3:exp_trace} shows a trace of conductance during the approach process of the tip towards the adatom directly below it, which is located  close to the Edge see Fig. \ref{fig2:sceneraio_STM}(b) or  atom A of  Fig. \ref{fig2:sceneraio_STM}(d)).
In Fig. \ref{fig3:exp_trace} the pale red area contains  values of trace of formation from the non-contact to the last point of tunnelling current ($G_t$). The red arrow indicates the direction in which trace of conductance should be read. The yellow area shows the region of the jump-to-contact(JC) that stretches from $G_{t}=0.05 G_{0}$ to atomic contact $G_{c}=0.95 G_{0}$, the orange arrow showing the abrupt jump in the conductance. The purple area is highlighting the region of the atomic contact.

We recorded eight traces of conductance in each scenario. From each trace we have extracted the pairs of points that represent the jump-to-contact, which are labeled as $G_t$ and $G_c$. Table \ref{Experimentaltable}  shows measured values of ($G_{t}, G_{c}$) in each scenario.
The precise preparation of the environment of the ad-atom sites and the reproducible tip  preparation procedure result in small variations of the values between the eight attempts. Nevertheless, some variation remains, as may be expected for a process that is associated with a dynamic instability.

\begin{table}[h!]
\caption{Jump-to-contact values  for adatoms on the \textcolor{black}{\textbf{Edge}} and \textcolor{red}{\textbf{Flat surface}}. The results of the jump-to-contact are expressed in pair of values ($G_{t}$, $G_{c}$) in units of $G_{0}$} 
\begin{tabular}{ c @{\quad} c @{\quad} c @{\quad} }
\toprule
\textrm{\textbf{Trace} }& \textrm{\textbf{Edge}} & \textrm{\textcolor{red}{\textbf{Flat surface}}} \\
\textrm{\textbf{Formation}}&\textrm{\textbf{$(G_{t}$, $G_{c}$)}} & \textrm{\textcolor{red}{\textbf{($G_{t}$, $G_{c}$)}}}\\
\colrule
01  & (0.17, 0.95) & \textcolor{red}{(0.04, 0.93)} \\
02  & (0.07, 0.94) & \textcolor{red}{(0.05, 0.93)} \\
03  & (0.05, 0.96) & \textcolor{red}{(0.04, 0.95)} \\
04  & (0.05, 0.96) & \textcolor{red}{(0.01, 0.91)} \\
05  & (0.06, 0.96) & \textcolor{red}{(0.02, 0.94)} \\
06  & (0.04, 0.95) & \textcolor{red}{(0.08, 0.94)} \\
07  & (0.03, 0.96)  & \textcolor{red}{(0.03, 0.94)} \\
08  & (0.08, 0.96) & \textcolor{red}{(0.03, 0.92)} \\
\botrule
\label{Experimentaltable}%
\end{tabular}
\end{table}

To summarize the results shown in table \ref{Experimentaltable} we have calculated the mean values and the standard deviations of $G_t$ and $G_c$ (see  table \ref{TableminorEXP}). We find that the last conductance before the jump into contact ($G_{t}$) to within our experimental accuracy does not depend on the positioning near the Edge. On the other hand, the first contact ($G_{c}$) values differ between the Edge and Flat surface scenarios.

\begin{table}[h!]%The best place to locate the table environment is directly after its first reference in text
\caption{Mean values and standard deviations of  $G_{t}$ and $G_{c}$ of the data shown in Table \ref{Experimentaltable}.}
\begin{tabular}{ c @{\quad}c @{\quad}  c @{\quad}}
\toprule
\textrm{Scenario}   &  \textrm{$G_{t}$}   &  \textrm{$G_{c}$} \\
\textrm{\enspace} &   \textrm{($G_{0}$)}  &  \textrm{($G_{0}$)} \\
\colrule
\textbf{Edge} & 0.07 $\pm$ 0.04 & 	0.96 $\pm$ 0.01 \\
\textbf{\textcolor{red}{Flat surface}} & \textcolor{red}{0.04} $\pm$  \textcolor{red}{0.02} &   \textcolor{red}{0.93} $\pm$  \textcolor{red}{0.01}\\
\botrule
\label{TableminorEXP}
\end{tabular}
\end{table}

\section{Theoretical Model}
The STM technique allows creating of different geometric scenarios and permits a study of the electronic transport, in order to assess whether or not bonding has occurred. On the other hand, classical molecular dynamics (CMD), gives us the ability to simulate the experiment atomistically, allowing us to estimate atomic structures and inter-atomic distances during adhesion, explaining the mechanisms underpinning the process of adhesion. The combination of density functional theory (DFT) and  non-equilibrium Green's function techniques (NEGF) allows us to determine conductance values for the structures obtained from CMD, permitting comparison with experimental conductance values for the respective contact scenarios, prior to the tips making contact.

\subsection{Classical Molecular Dynamics simulations}
In order to imitate the experiments, we have simulated the two different scenarios represented by Figs. \ref{fig2:sceneraio_STM} a) and b), respectively. Thus, panel a) shows an atomically sharp gold tip directly above an adatom on top of an otherwise smooth surface and b), the same tip above an adatom close to a step edge. Each also represents the initial structures used in the respective simulations,  prior to the tips making contact. 

Classical molecular dynamics (CMD) is based on solving the second Newton's law  to obtain the trajectory of every atom during a simulation \cite{allen1989computer,frenkel2002}. In this way, we can trace the evolution of a collection of atoms in real-time, and at a very fine time resolution ($\sim 1.0$ femtosecond in our case). The interaction between the atoms in the simulation is usually derived semi-empirically, with the many-body potential being fitted to a number of physical properties of the material, obtained experimentally and/or by means of first-principles quantum mechanical calculations. Here, we use the semi-empirical many-body embedded atom method (EAM) \cite{daw1983semiempirical}  with the interatomic potential described by Zhou \textit{et al.} \cite{zhou2001atomic,wadley2001potential}. This potential reproduces the elastic and mechanical properties of atomic-sized gold contacts surprisingly well. \cite{Fernandez2016,Sabater2018}

In  this work, we use the \texttt{LAMMPS} \cite{plimpton1995fast,lammps2} CMD code to model a gold tip interacting with a gold surface oriented along the [111] crystallographic direction (see Fig. \ref{fig2:sceneraio_STM}), to generate these tip-adatom-surface structures. The adatom on a Flat surface,  Fig. \ref{fig2:sceneraio_STM}(a), or at an Edge, Fig. \ref{fig2:sceneraio_STM}(b), contain a total of 1830 and 1746 atoms, respectively. To move the tip towards the respective adatoms, we freeze the internal positions of the atoms in the topmost layers of the tip, which are then moved downwards at a speed of $\sim 1.0$ m/s. The lowest layers of the substrate gold surface are also frozen, and are kept in place with no bulk motion. The remaining atoms of the tip-surface system respond dynamically to the motion of these frozen layers and, in this way, tip and surface (with adatom on top) are pushed together in order to make contact. The surfaces are periodic in the \textit{x}- and \textit{y}-directions. The speed with which the approach occurs may be several orders of magnitude higher than in an actual experiment, but it is still low enough for the atoms to reach equilibrium between thermostating to the target temperature, which is effected every 1000 time steps by means of the Nose-Hoover thermostat \cite{nose1984molecular,hoover1985canonical}. This particular thermostat was chosen since it reproduces the canonical ensemble even in the presence of an external force such as that applied to move our tip \cite{frenkel2002}. For integration of the trajectory of the system, the Velocity-Verlet algorithm was used \cite{swope1982,frenkel2002}, with an integration time step of 1 fs.

In order to determine when first-contact occurs, we have established the next criteria. First of all, in our simulations we record the distance between the apex atom of the tip and the adatom. We assumed  that the first contact occurs when these two atoms are within a distance of each other that is halfway between first and second nearest neighbors in a perfect FCC lattice. In the case of gold this corresponds to $3.48$ \AA. Once contact is detected the first point of contact and the last point of the approach can be identified. In order to compare this with the corresponding experiment, we need to calculate electronic transport for the  scenarios of pre-contact and first point of contact.

\subsection{Electronic transport calculations}
In this manuscript, we have computed the electronic transport using  DFT+NEGF. We use the non-equilibrium Green's function (NEGF) approach to quantum scattering, in particular, the Keldysh formulation thereof \cite{palacios2001fullerene,palacios2002transport,louis2003keldysh}. This method has been implemented in the Atomistic Nano Transport (\texttt{ANT.G}) code \cite{ANTG}, which interfaces with the quantum chemistry DFT code \texttt{Gaussian} \cite{GAUSSIAN09}. \texttt{ANT.G} divides the nanocontact into a semi-infinite lead - device - semi-infinite lead configuration. \texttt{Gaussian} then calculates, within the  local spin density approximation  of the exchange-correlation function of DFT, the electronic structure of the device region, including a portion of the semi-infinite leads on both sides of the device. The extended leads (constructed as Bethe lattices) are, in turn, described electronically by Slater-Koster tight-binding parameters \cite{palacios2002transport}. The calculations would become costly if all the atoms in a given CMD snapshot were included. Therefore, before performing conductance calculations, we trim down our CMD snapshots to fewer than 400 atoms. The trimmed-down structures are centered on the region of first contact. Furthermore, to obtain accurate results, we assign an 11-electron \textit{spd}-orbital basis set to around 100 atoms in the constriction \cite{Dednam2015,Fernandez2016}. The remaining atoms are assigned a one-electron 6\textit{s}-orbital basis set.

In order to imitate the experimental approach of continuous cycles of repeated contact-formation and breaking, the CMD simulations are similarly performed. A total of 10 continuous cycles of rupture and formation were simulated for each scenario of the adatom (near the Edge or on a Flat surface, respectively). The conductance for the tunneling $G_t$ and contact $G_c$ regimes are then determined using the above approach, for each of the cycles, for each of the two adatom scenarios.

\section{Theoretical Results}
\begin{figure}[ht!]
 \centering
 \includegraphics[width=0.5\textwidth]{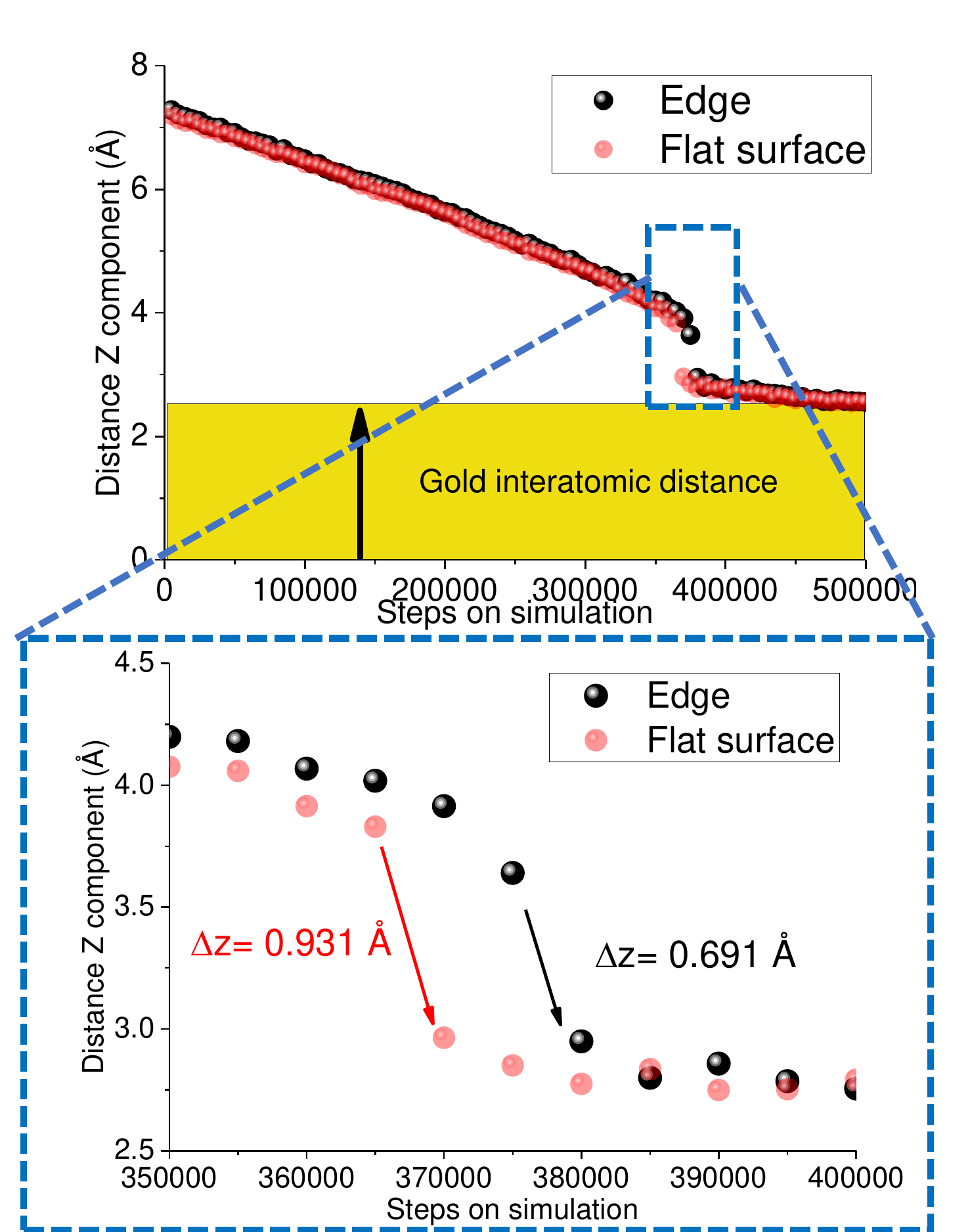} 
 \caption{Vertical distance between tip apex atom and adatom vs simulation time step. Black dots correspond to the adatom on a Edge scenario, and red dots to adatoms close to the Flat surface. Inset: A zoom-in of the distance traces near the moment when contact first occurs.}
 \label{fig4:MDresults111}
\end{figure}

Applying the algorithm described above, we repeat the contact formation process for ten cycles for each configuration in the CMD simulations. Mirroring the experimental approach, we can represent the vertical distance between the adatom and tip apex atom vs. simulation time step, as shown in  Fig. \ref{fig4:MDresults111}. Both the Flat surface and Edge scenarios are shown, with light red dots corresponding to the adatom in the Flat surface region, and  black dots for adatoms located in the first hollow site on top of the atomic step. For sake of comparison with the experiment, we refer to this type of plot as a distance trace. In Fig.~\ref{fig4:MDresults111}~the traces of distance correspond to cycle number 5 in both scenarios. The dashed rectangle in the upper panel is magnified in the bottom panel to show more clearly how the simulated jump-to-contact occurs. The jump seems to occur less abruptly when the adatom is near an Edge, while, when the adatom is on the Flat surface, the jump distance is larger and more abrupt.

Upon analyzing all the calculated distance traces, we obtain the vertical atomic distance in contact (contact distance), the distance prior to contact (tunnel distance), and the difference between these two distances $\Delta z$. Tables \ref{EDGE-TABLE} and \ref{FLAT-TABLE} show these CMD results in the first four columns, for the Edge and Flat surface scenarios, respectively. The first column corresponds to the contact cycle, the second and third columns show respectively the distance (in \AA) between the adatom and the apex atom of the tip in the tunneling and contact regimes. The fourth column shows the jump distance $\Delta z$ in going from tunneling to contact. Finally, the last two columns contain the results of conductance calculations in units of $G_{0}$ obtained via DFT+NEGF transport for tunneling and contact regime, respectively.

 \begin{table}[h!]
%\caption{Molecular dynamics and DFT results for the Edge and Flat surface scenarios}
\caption{Molecular dynamics and conductance calculations via DFT+NEGF for the {\textbf{Edge}} scenario }
\begin{tabular}{c@{\quad}  c@{\quad} c@{\quad} c@{\quad}   |c@{\quad}  c@{\quad} c@{\quad}}
\toprule
Cycle & $z$ tun. & $z$ con. & $\Delta$z & $G_t$ & $G_c$\\
    & (\AA) & (\AA) & (\AA) & ($G_0$) & ($G_0$) \\
\colrule
01  &  3.74 & 2.87 & 0.87 &  0.30 & 0.94 \\
02  &  3.64 & 2.93 & 0.70 & 0.32 & 0.83\\
03  &  3.65 & 2.94 & 0.71 & 0.38 & 0.86\\
04  &  3.66 & 2.92 & 0.75 & 0.34 & 0.86\\
05  &  3.64 & 2.95 & 0.69 & 0.36 & 0.85\\
06  &  3.62 & 2.93 & 0.69 & 0.35 & 0.87\\
07  &  3.53 & 2.92 & 0.61 & 0.39 & 0.84 \\
08  &  3.77 & 3.31 & 0.47 & 0.27 & 0.58\\
09  &  3.60 & 2.95 & 0.65 & 0.36 & 0.82\\
10  &  3.53 & 2.90 & 0.62 & 0.45 & 0.83\\
\colrule
%\botrule
\label{EDGE-TABLE}%
\end{tabular}
\end{table}

 \begin{table}[h!]
%\caption{Molecular dynamics and DFT results for the Edge and Flat surface scenarios}
\caption{Molecular dynamics and conductance calculations via DFT+NEGF for  \textcolor{red}{\textbf{Flat surface}} scenario}
\begin{tabular}{c@{\quad}  c@{\quad} c@{\quad} c@{\quad}   |c@{\quad}  c@{\quad} c@{\quad}}
\toprule
Cycle & $z$ tun. & $z$ con. & $\Delta$z & $G_t$ & $G_c$\\
    & (\AA) & (\AA) & (\AA) & ($G_0$) & ($G_0$) \\
\colrule
\textcolor{red}{01}  & \textcolor{red}{3.71} &  \textcolor{red}{2.97} &  \textcolor{red}{0.74} &\textcolor{red}{0.23} & \textcolor{red}{0.84}\\
\textcolor{red}{02}  & \textcolor{red}{3.77} &  \textcolor{red}{2.98} &  \textcolor{red}{0.79} &\textcolor{red}{0.28} & \textcolor{red}{0.81}\\
\textcolor{red}{03}  & \textcolor{red}{3.89} &  \textcolor{red}{2.88} &  \textcolor{red}{1.01} &\textcolor{red}{0.23} & \textcolor{red}{0.92}\\
\textcolor{red}{04} &  \textcolor{red}{3.78} &  \textcolor{red}{2.89} &  \textcolor{red}{0.89} &\textcolor{red}{0.32} & \textcolor{red}{0.87}\\
\textcolor{red}{05}  & \textcolor{red}{3.89} &  \textcolor{red}{2.96} &  \textcolor{red}{0.93} &\textcolor{red}{0.27} & \textcolor{red}{0.87}\\
\textcolor{red}{06} & \textcolor{red}{3.82} &  \textcolor{red}{2.76} &  \textcolor{red}{1.06} &\textcolor{red}{0.24} & \textcolor{red}{0.93}\\
\textcolor{red}{07} & \textcolor{red}{3.76} &  \textcolor{red}{2.80} &  \textcolor{red}{0.96} &\textcolor{red}{0.34} & \textcolor{red}{0.95}\\
\textcolor{red}{08} &\textcolor{red}{3.76} &  \textcolor{red}{2.84} &  \textcolor{red}{0.92} &\textcolor{red}{0.34} & \textcolor{red}{0.94}\\
\textcolor{red}{09} & \textcolor{red}{3.80} &  \textcolor{red}{2.89} &  \textcolor{red}{0.91} &\textcolor{red}{0.30} & \textcolor{red}{0.89}\\
\textcolor{red}{10} & \textcolor{red}{3.76} &  \textcolor{red}{3.05} &  \textcolor{red}{0.71} &\textcolor{red}{0.33} & \textcolor{red}{0.84}\\
\botrule

\label{FLAT-TABLE}%
\end{tabular}
\end{table}

From Table~\ref{EDGE-TABLE}, the mean distance between the apex atom and the adatom at the jump to contact, averaged over the ten cycles, is $3.64 \pm 0.08$ \AA, while after contact has been established, it is $2.96 \pm 0.12$ \AA, with the mean jump distance $0.68 \pm 0.1$ \AA. For the 10 cycles for the Flat surface, the mean distance before the jump is $3.79 \pm 0.06$ \AA, after contact it is $2.89 \pm 0.08$ \AA, and the mean jump distance is $0.89 \pm 0.11$ \AA . We summarize in Table~\ref{Tableminor} these averaged results from Tables~ \ref{EDGE-TABLE} and \ref{FLAT-TABLE}. The first column indicates the scenario, Edge or Flat surface. The second column shows the mean value and standard deviation of $\Delta$z, and the third and fourth columns show the mean value and standard deviation of $G_{t}$ and $G_{c}$ respectively. 

\begin{table}[h!]
\caption{Mean values and standard deviations of $\Delta$z, $G_{t}$ and $G_{c}$ of the data shown in Tables \ref{EDGE-TABLE} and \ref{FLAT-TABLE}.}
\begin{tabular}{ c @{\quad}c @{\quad} c @{\quad} c @{\quad}}
\toprule
Scenario  & $\Delta z$ &  $G_t$   &  $G_c$ \\
\enspace & (\AA)   &  ($G_0$)  &  ($G_0$) \\
\colrule
\textbf{Edge} & 0.68 $\pm$ 0.10 & 0.35 $\pm$ 0.05 & 	0.83 $\pm$ 0.09 \\
\textbf{\textcolor{red}{Flat surface}}& \textcolor{red}{0.89} $\pm$ \textcolor{red}{0.11} & \textcolor{red}{0.29} $\pm$  \textcolor{red}{0.04} &   \textcolor{red}{0.89} $\pm$  \textcolor{red}{0.05}\\
\botrule
\label{Tableminor}
\end{tabular}
\end{table}

\section{Discussion}

In this section we will discuss how the atomic level detail of the mechanical properties of the environment at the junction can effect the dynamic bonding behaviour. 

\subsection{Effect of variations in the surface structure}
\begin{figure}[ht!]
\centering

\includegraphics[width=0.5\textwidth]{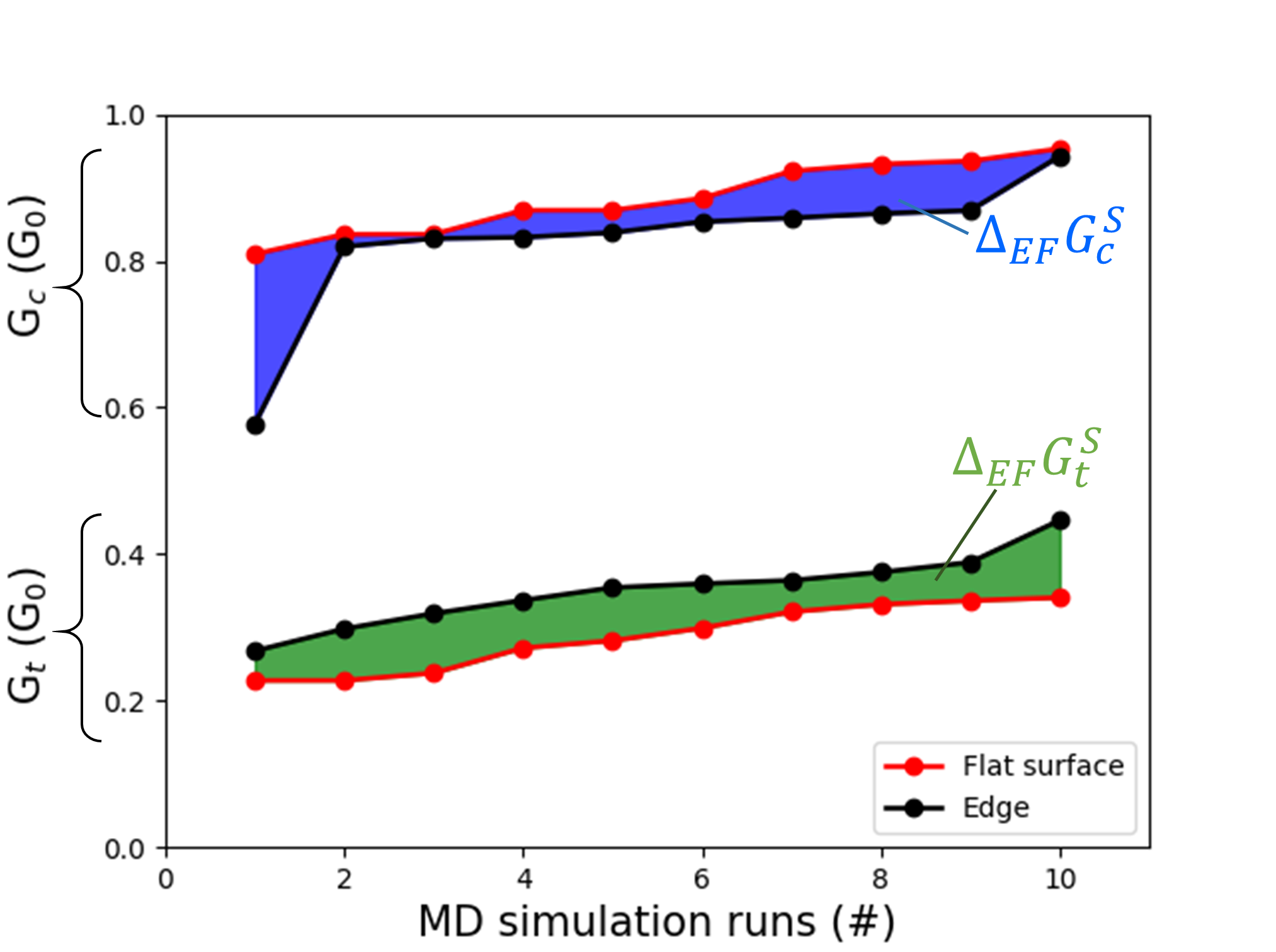}
\caption{Line plots showing sorted $G_{t}$ and $G_{c}$ values of calculated conductance for Flat surface (red) and Step Edge (black). Blue shows the difference($\Delta$) between calculated conductance of the contact ($G_{c}^{S}$) for the Flat surface and Edge (F and E sub-index). Green region correspond on the difference difference($\Delta$) between calculated conductance of the tunneling regime ($G_{t}^{S}$)for the Flat surface and Edge.}
\label{fig5:DFT_MDa_expb}
\end{figure}

To assess how the change in the surface affects the dynamic bonding, we have studied two scenarios as mentioned earlier. In both the scenarios a STM tip is made to approach from above an adatom placed on top of (1) a Flat Au (111) surface and (2) a monoatomic step Edge created over a Au(111) surface. All the corresponding simulation cycles (as discussed in the previous section) were performed using the same ideal crystalline pyramidal tip structure. This helps in avoiding any effect of variations in tip structures at this moment, and, as a result, the variation in the simulation results comes entirely from the small temperature fluctuations. Therefore, because the temperature fluctuations are completely random in our ergodic simulations within the canonical-ensemble \cite{nose1984molecular,hoover1985canonical,frenkel2002}, we can, for convenience, sort the $G_{t}$ and $G_{c}$ values given in Tables \ref{EDGE-TABLE} and \ref{FLAT-TABLE} in ascending order. We can then plot a combined line plot with conductance on the vertical axis and count index (ranging from 1 to 9) on the horizontal axis for both the scenarios (black for Edge and red for Flat surface), see Fig.~\ref{fig5:DFT_MDa_expb}.

Comparing the results  for the Flat surface and Edge case of Fig.~\ref{fig5:DFT_MDa_expb}, we observe that the mean value of $G_t$ is higher in the case of the Edge. The mean $G_c$ values are however very similar in both the scenarios, with marginally higher numbers for the Flat surface. The marginally higher $G_c$ values for the Flat surface can be understood by higher coordination value \cite{Sabater2018} of the point contact in case of the Flat surface. A similar state of affairs is expected for the $G_{t}$ values, which should exhibit a slightly lower value in the case of the simulated step Edge. However, what  we observe in our simulations is rather the opposite. We show a similar conclusion derived also from experiments in the next section.
Up to this point, we cannot discern whether it is a consequence of the geometrical configuration or attributable to a difference in the density of states.

To understand the origin of the high values of $G_{t}$ in the Edge scenario, we consider two main hypotheses. The first is based on the Smoluchowski effect \cite{ibach2006physics}, an electronic effect in which the adatom is a tightly bound surface dipole, leading to the adatom being held more strongly on the Flat surface than at the Edge, with lower tunneling conductance as a result. 

The second hypothesis is based on a geometric effect and mechanical surface properties, in which the tip pulls on the adatom as it approaches it, with the adatom in turn pulling on the atoms of the step Edge, bending the surface towards the tip. To explore both hypotheses, we have selected the following atomic geometry configurations, cycle 3 and 5 for Edge case and Flat surface, respectively. We have selected these cycles since the $G_c$ values are quite similar but the $G_t$ is almost twice as high in the case of the Edge than the Flat surface.

To explore the role of the Smoluchowski effect, we have performed total energy calculations in \texttt{Quantum ESPRESSO} \cite{giannozzi2009QE} on the Edge and Flat surface structures from our CMD simulations, excluding the tips. Based on the difficulty of calculating reliable surface states using (unreconstructed) surfaces obtained from CMD and dealing properly with periodic boundary conditions for large super cells in DFT calculations, we have decided to base our calculations on an unreconstructed (001) surface. However, the results should be similar for a unreconstructed (111) surface, since the purpose is to illustrate qualitatively how the Smoluchowski effect might come into play in these two scenarios. Therefore, as in most of our quantum transport calculations, we did not further relax  the atoms and here simply used the respective atomic structure snapshots from the CMD simulations, at the point directly prior to jumping into contact, and directly after, and calculated the total energy of the CMD snapshots with periodic boundary conditions in the $x$ and $y$ directions. We used the Perdew-Zunger LDA exchange-correlation functional \cite{perdewzung1981} and an ultrasoft pseudopotential that has been benchmarked against all-electron calculations \cite{Lejaeghere_2016}. We also verified the convergence of the total energy as a function of $k$-point sampling in the irreducible wedge of the Brillouin zone. A $2 \times 2 \times 1$ Monkhorst-Pack mesh provided good convergence given the size of the supercell ($16.32\times16.32\times24.0\/ $\,\AA$^3$).

\begin{figure}[h!]
\centering
\includegraphics[trim={2cm 2cm 0 0},clip,width=0.49\textwidth]{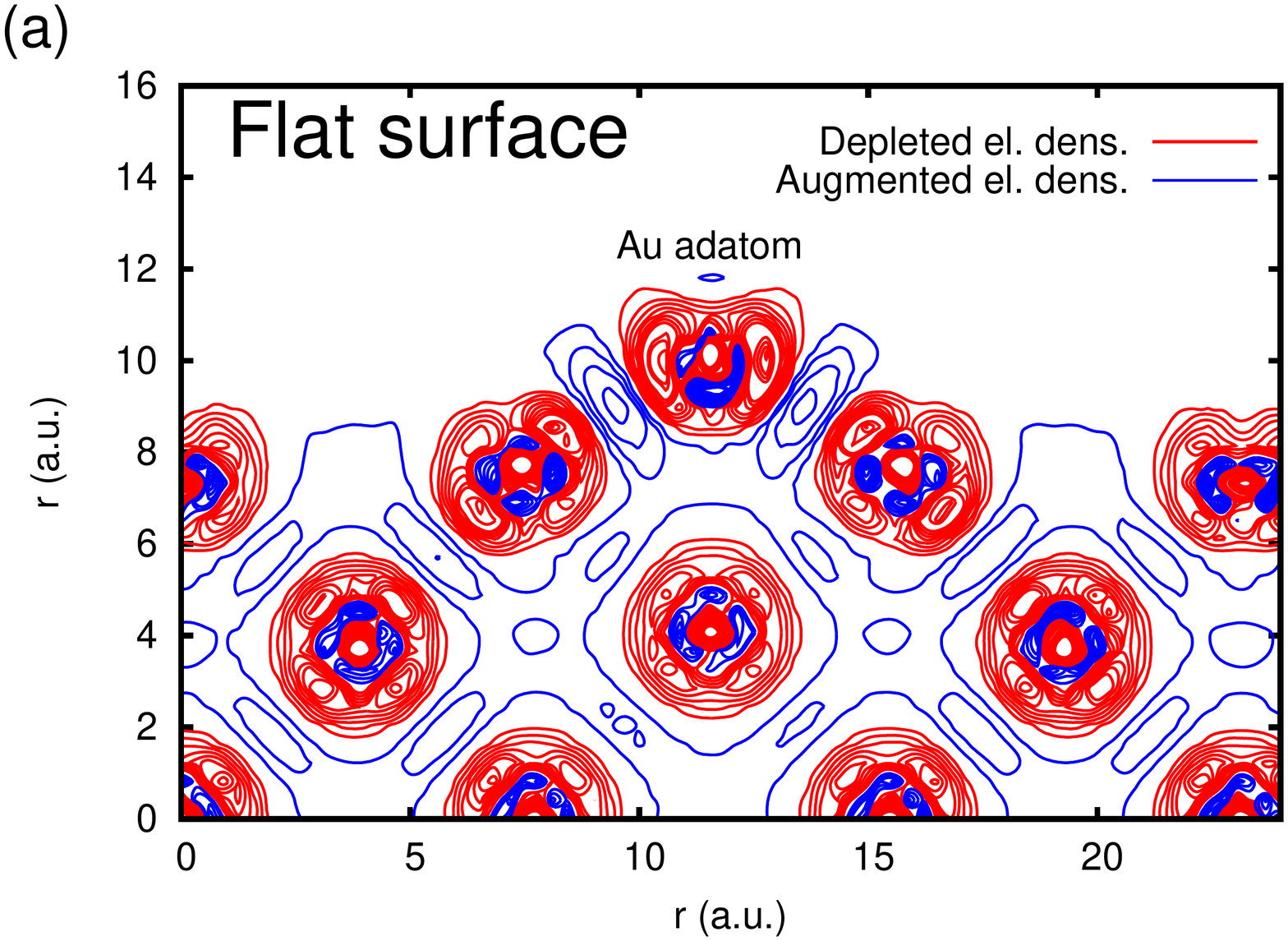} 
\includegraphics[trim={2cm 2cm 0 0},clip,width=0.49\textwidth]{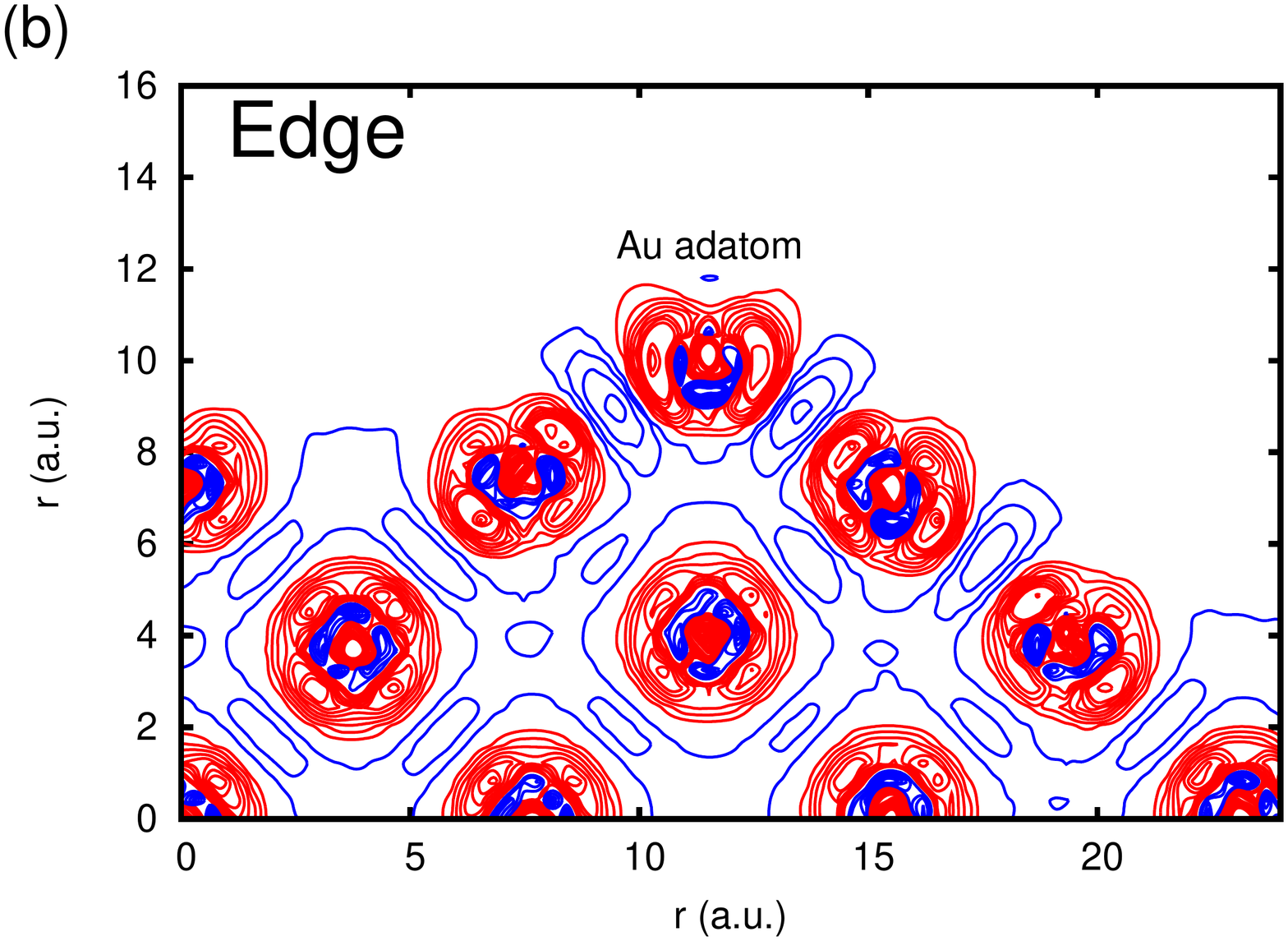} 
\caption{Electronic density difference contours on planes passing through the center of an adatom on a), a gold Flat surface and b), near a step Edge. Red and blue contours correspond to negative and positive values of the electron density difference, representing regions depleted of and augmented with electron density relative to free Au atoms, respectively.} 
\label{fig6:B05E03Smoluchowski}
\end{figure}

In Fig. \ref{fig6:B05E03Smoluchowski}, we thus show electronic density difference contours on transverse planes through the center of the adatom on the adatom-surface systems. Panel a) corresponds to the case of a Flat surface and panel b) to the adatom near an Edge. Moreover, Fig. \ref{fig6:B05E03Smoluchowski} shows in both cases that the adatom shares significantly more electron density with its neighbours in the fourfold (001) hollow (see the blue contour lines), than for the other atoms on the surface. The role of the Smoluchowski effect \cite{ibach2006physics} therefore is similar in both cases, and hence cannot explain the observed differences between mean $G_{t}$ values in Edge and Flat surface.

\begin{figure}[!ht]
 \centering
   \includegraphics[width=0.49\textwidth]{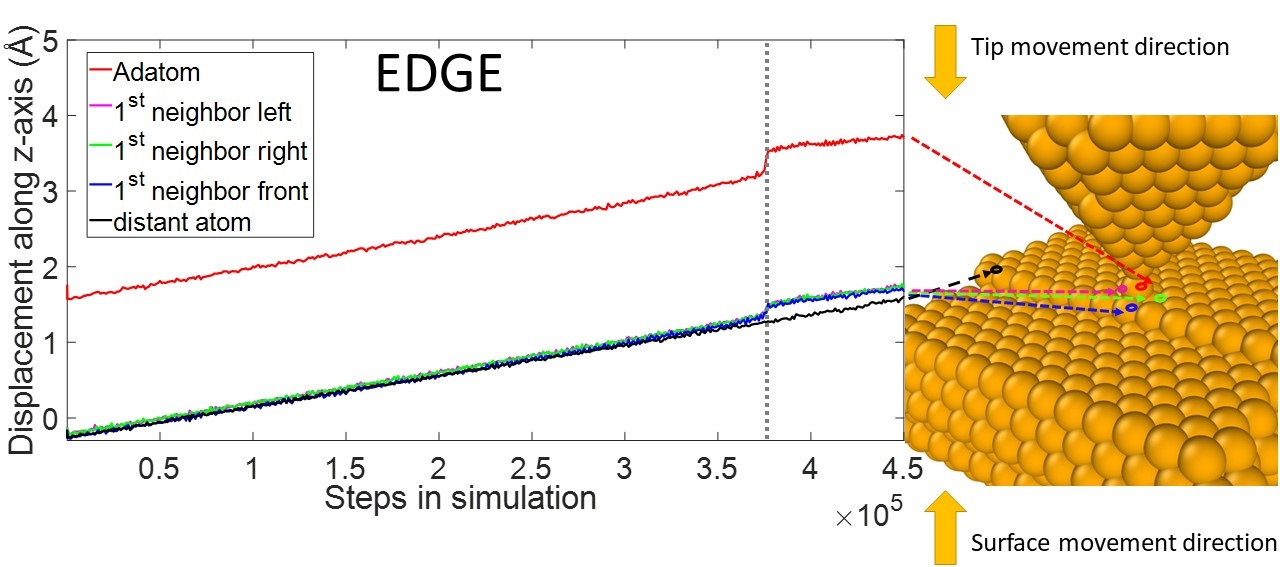}
   \includegraphics[width=0.49\textwidth]{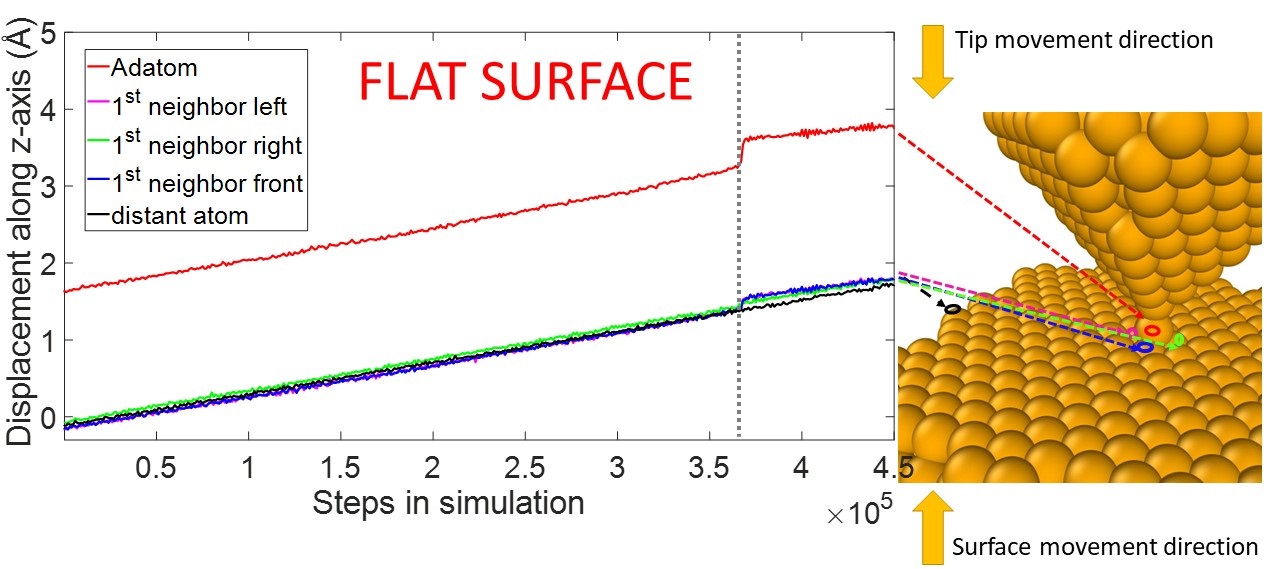}
   \caption{Top and bottom panels correspond to the trajectories along the $z$-axis of selected atoms in the depictions of the Edge and Flat surface shown on the right in the insets. Red, pink, green, blue and black lines are the trajectories of the adatom as well as its first neighbors to the right, left, front and a distant atom.}
   \label{fig7:neigZstudy}
 \end{figure}

After discarding the Smoluchowski effect, we explore the second hypothesis. We analyze the deformations over the surface that occurs when the adatom is attracted by the apex atoms for the simulated cycles 3 and 5.
In this analysis, we plot the trajectory along the $z$-axis of the adatom and its first neighbors to the right, left, front and an atom that is distant from the point contact. Figure \ref{fig7:neigZstudy} shows the trajectories of these atoms using different colors, together with that of a distant atom from the same row of atoms.

Moreover, we have selected a distant atom from the same row of atoms, which has been color-coded black. The upper panel in Fig. \ref{fig7:neigZstudy} shows the trajectories for the Edge scenario and the bottom panel corresponds to the Flat-surface case. The insets to the right of each panel of trajectories depict the two scenarios. Dashed lines guide the eye to identify the atoms corresponding to the trajectories. Furthermore, the yellow arrows indicate the relative movement of the two surfaces and tips. According to the movement  of the surface all the trajectories are ascending slopes. For all the first neighbors as well as the adatom trajectories, we observe a jump in the trend of the slope. This jump occurs when the adatom jumps to make contact with the apex atom of the tip.  That the neighbor atoms also jump confirms that the surfaces are deformed by the forces that pull on the adatom. On the other hand, we observe that in both cases the distant atom does not suffer such a jump and, as a consequence, its adjoining borders are not deformed.
 
\begin{figure}[!ht]
 \centering
   \includegraphics[width=0.5\textwidth]{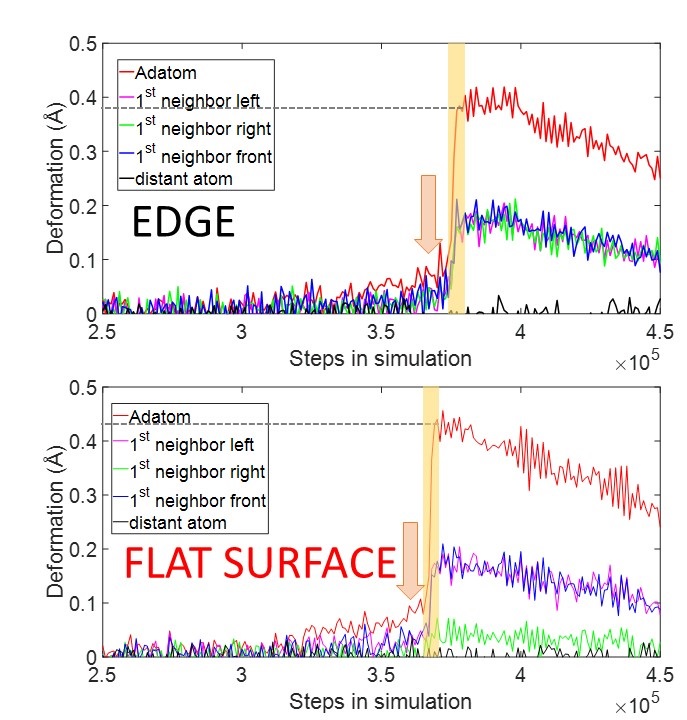}
   \caption{Top and bottom panels correspond to the deformation of the trajectory calculated from the subtraction of the slope of the trajectory from the trajectory itself. Upper panel corresponds to Edge and bottom to Flat surface scenario. The colors of the traces refer to the same atoms as in the previous figure.}
 \label{fig8:deformation}
\end{figure}

However, based only on these trajectories it is difficult to draw a definite conclusion. Therefore, to better understand the  mechanical effects, we have studied the normalized deformation of the trajectory, which we have obtained by subtracting the slope of every trajectory from the trajectories themselves, as Fig. \ref{fig8:deformation} shows. Within this analysis, the upper and lower panels show the deformation of the Edge and Flat surface, respectively. The colored lines refer to the same atoms as in Fig. \ref{fig7:neigZstudy}. The vertical yellow highlighted areas brings attention to the abrupt deformation that occurs in all the curves. Figure  \ref{fig8:deformation} shows that for the Edge case more steps of simulations are required to produce the abrupt deformation (``the jump-to-contact''). Moreover, in both panels, the distant atom (black traces in Fig.~\ref{fig8:deformation}) does not exhibit a deformation over the trend in its trajectory, as expected. Considering the deformation of the three first neighbors of the adatom, to the front (blue), left (pink) and right (green), we see that in the case of an adatom at the Edge, all first neighbors undergo a similar deformation during the jump to contact. In contrast, for the adatom in the the Flat surface region, while the front and left adatoms undergo similar deformations as in the Edge case, the right neighbor in the Flat case (green) undergoes a much smaller deformation ($\sim$4 times smaller) than in the Edge case.  This implies that the right neighbor will not contribute to the tunneling current to the same extent in the Flat surface case as in the Edge case \cite{Sabater2018}.  Therefore, while all three neighbors fully contribute to the tunneling current for the Edge case, only two neighbors will substantially contribute to the tunneling current in the Flat surface case. This explains why the mean value of $G_t$ is lower for the Flat surface than the Edge case.

Another characteristic of Fig. \ref{fig8:deformation} is that the maximum value of the deformation is larger for the Flat surface than for the Edge (see grey dashed lines). This fact explains why the distance of the jump to contact ($\Delta z$) is higher in the Flat surface than in the Edge. In other words, this large displacement of the adatom on the Flat surface occurs because before the jump to contact, the adatom is strongly anchored to the surface thanks to larger number  of second neighbor atoms that this surface has compared to the Edge.

An additional striking feature of Fig. \ref{fig8:deformation} is marked by the arrow, which indicates where the deformation preceding the jump-to-contact commences. In the case of the Edge all the deformation curves of the first neighbors follow the deformation trend of the adatom, increasing smoothly before the abrupt jump occurs. However in the case of the Flat-surface the adatom starts to deform significantly earlier than the neighboring atoms, which only start to deform once the abrupt jump-to-contact occurs. An immediate conclusion is that on the Flat surface the adatom initiates a deformation earlier and its first neighbors are not affected in the early stages of this process; they are only affected a few steps before the jump to contact of the adatom occurs.
This transition is a consequence of the fact that the Flat surface deforms less readily than the Edge. Recall that we found the strength of the Smoluchowski effect to be similar in both situations, and that the adatom is therefore bound nearly equally strongly on both surface types. So, early in the contact making phase, unlike on the Edge, the adatom on the Flat surface is not accompanied by its nearest neighbours during the deformation process.
It is clear that the greater relative freedom of the adatom in the Flat surface case allows this adatom to approach the tip apex more closely than for the Edge case. It is for this reason that the calculated conductance $G_t$ in the table \ref{Tableminor} is lower in the case of the Flat surface. Note also that all the calculated tunneling conductance values are an order of magnitude larger than in the experiments because the CMD simulations cannot fully account for the particularly strong force between gold atoms from scalar relativistic effects \cite{Calvo2018}

\subsection{Effect of variations in experimental STM tip structure}

In the theoretical models discussed in the previous section, we have assumed a well defined pyramidal STM tip and have shown how small changes in the mechanical deformations on the surface have significant impact on the dynamic bonding and associated conduction measurements. However, experimentally, the structure of the STM tips is usually unknown and is always a matter of concern in any atomic scale electronic transport measurements \cite{Ferdinand2020}. There are techniques to train the shape of the tip apex up to one or two atomic layers \cite{sabater2012mechanical,Sumit2017b}, and is usually assumed that as the tunneling current drops exponentially with distance, the tip structure behind the apex should not play a significant role. 

In order to compare the \textit{ab-initio} data shown in Fig.~\ref{fig5:DFT_MDa_expb} in the two scenarios with the experiment, we have studied the difference between experimental and calculated mean values of $G_{t}$ and $G_{c}$.

\begin{figure}[ht!]
\centering
\includegraphics[width=0.5\textwidth]{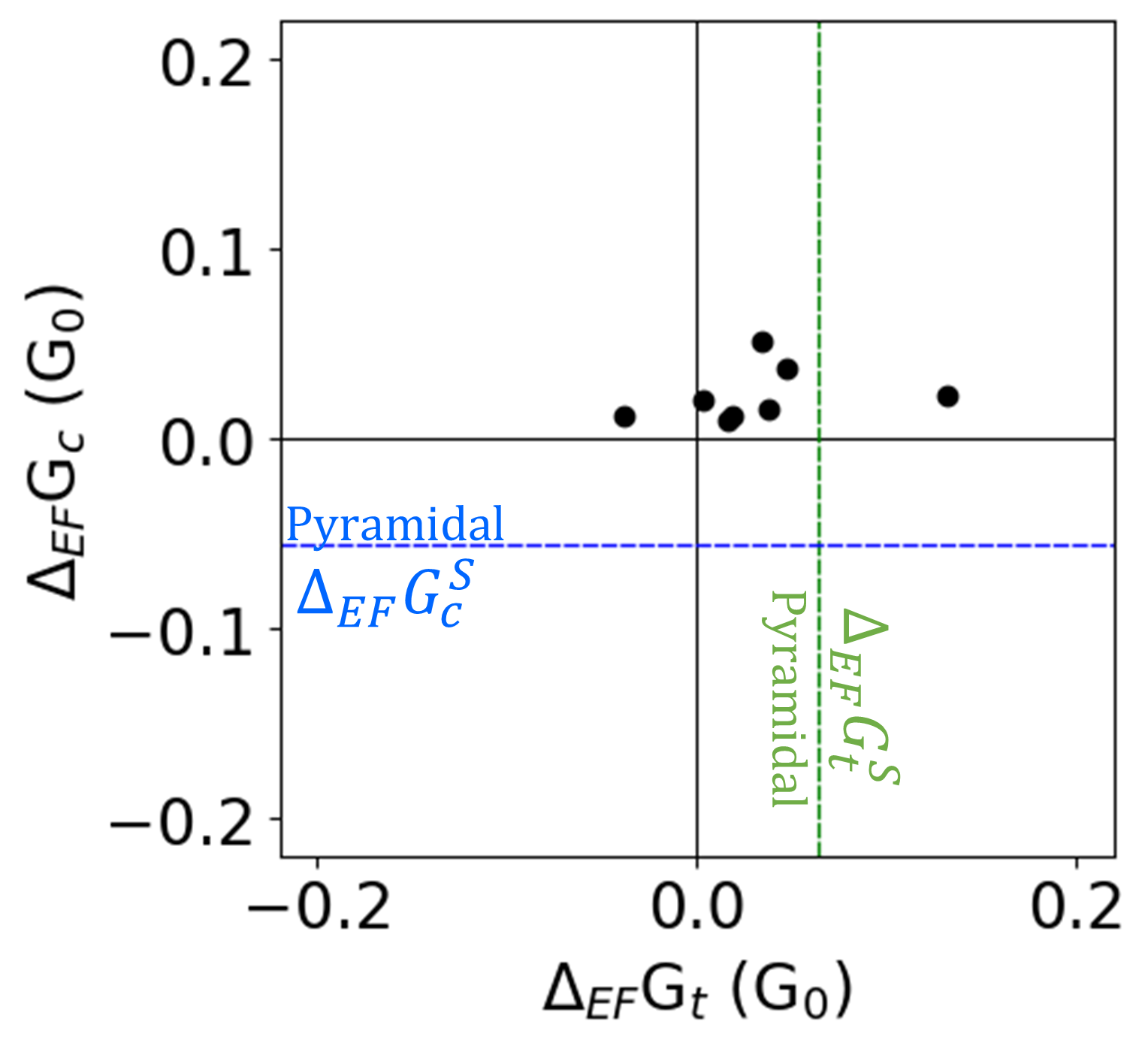}
\caption{Experimental data of $\Delta_{EF}G_{c}$ vs. $\Delta_{EF}G_{t}$. Green and blue dashed lines mark the position of \textcolor{ForestGreen}{$\Delta_{EF}G_{t}^{S}$} and \textcolor{RoyalBlue}{$\Delta_{EF}G_{c}^{S}$} values extracted for an ideal pyramidal tip from \textit{ab inito} calculations.}
\label{fig9:DFT_MDa_exp}
\end{figure}

In Table \ref{Experimentaltable},  we collect the conductance data of first contact and last value of tunnelling, extracted from 8 experimental traces of conductance, where each set consists of measurement on top of the Flat-surface and then on top of the step Edge. During each set of experiments comprising a measurement at the Edge and a measurement at the Flat surface, we avoided changes in the tip apex structure \cite{Sumit2017b}. Then the tip was retrained by means of the mechanical annealing technique \cite{sabater2012mechanical,Sumit2017b} before the next set of experiments were conducted. To compare the experimental results with our calculations, we have defined a parameter $\Delta_{EF}$, which is the difference between the experimental conductance's values of Edge and the Flat surface scenarios within each set \textit{i.e.,} $\Delta_{EF}=G_{\text{Edge}} - G_{\text{Flat Surface}}$. $\Delta_{EF}$ values are calculated separately for the tunnelling ($\Delta_{EF}G_{t}$) and first contact conductance values ($\Delta_{EF}G_{c}$).  These are then plotted in a $\Delta_{EF}G_{c}$  vs. $\Delta_{EF}G_{t}$ plot in Fig. \ref{fig9:DFT_MDa_exp}. The black dots represent the experimental data points. We have also plotted blue and green dashed lines which represent the mean $\Delta_{EF}$ values obtained from the simulation data shown in Fig. \ref{fig5:DFT_MDa_expb}. These are called \textcolor{RoyalBlue}{$\Delta_{EF}G_{c}^{S}$} and \textcolor{ForestGreen}{$\Delta_{EF}G_{t}^{S}$}, where the superscript 'S' signifies that these are obtained from the simulated data using a pyramidal tip.

Considering that the tip apex structure in all the experiments can be assumed similar to a one or two layer pyramidal structure, we would expect that the $\Delta_{EF}G_{t}$ and $\Delta_{EF}G_{c}$ values would be comparable to those from the theoretical simulations. Indeed, we do see some similarities, like the $\Delta_{EF}G_{t}$ values obtained experimentally are mostly positive similar to the simulation (green dashed line) and which could be explained by the extra bulging of the surface at the step Edge also discussed earlier. However, the experimental values still differ significantly from the simulation. In this regard, we highlight that our calculated $G_{c}$ values are underestimated by $\sim 5-10\%$ because our method does not account for surface states \cite{Li2021}, the implementation of which we leave for future work. Nevertheless, surface states are expected to be present in both cases, though should be less important for the Edge. So, in calculating \textcolor{RoyalBlue}{$\Delta_{EF}G_{c}^{S}$}, surface state effects should cancel out to some degree. After accounting for the limitations of our calculations, our combined theoretical and experimental results suggest that although in our controlled experiments the tip apex structure was trained, the experimental tip would be very different from the ideal pyramidal structure far from the tip apex. This would confer different mechanical properties to the experimental tips and they would thus behave differently in the dynamic bonding experiments.

Such an analysis therefore also helps to establish whether an experimental tip would be closer to an ideal crystalline pyramidal tip. So, with such controlled tests we can get an idea that the structure of the experimental tips for which the $\Delta_{EF}G_{t}$ and $\Delta_{EF}G_{c}$ values would fall close to the simulated values (\textcolor{RoyalBlue}{$\Delta_{EF}G_{c}^{S}$} and \textcolor{ForestGreen}{$\Delta_{EF}G_{t}^{S}$}), should be then closer to the ideal pyramidal tip structure even far from the tip apex or at least shares similar mechanical properties.

\section{Conclusions}
In this work, we have shown that repeated approach cycles of atomic sized tips on adatoms either close to a monatomic step edge, or in an atomically flat surface region far from any step edge on Au (111) facets, exhibit dispersion in both theoretical and experimental conductance data. This fact reveals that small changes in the surroundings of the adatom and tip affect the conductance during the pre- and first contact processes.

We have discussed here two scenarios. First where an adatom is placed over a flat Au(111) surface was approached from above using an STM tip and second where the adatom was placed at a monoatomic step-edge and the same experiment was repeated. We show here using both experiment and simulation that tunneling conductance ($G_t$) just before the jump-to-contact is larger in the case of the step edge, which could be counter intuitive.

We explained the origin of the difference in $G_t$ values under the two scenarios via a hypothesis involving the role of the geometry of the surface.

Furthermore, we show that as the ``dynamic bonding'' behavior is sensitive to fine adjustments in the mechanical properties of the junction, we can use it to extract insights about the STM tip beyond just apex of the tip. Finally, the demonstrated sensitivity of the dynamical bonding of the STM tip (and the associated conductances) as a function of deformation and local environment at the atomic level, can contribute to the advancement and improve understanding of the interaction of STM tips with atomic-scale structures in the emerging field of straintronics.

\section{Acknowledgment}
We gratefully acknowledge the generosity and useful discussions of Professor M.J. Caturla and C. Untiedt from the University of Alicante.
This work was supported by the Generalitat Valenciana through CDEIGENT2018/028, PROMETEO2017/139, PROMETEO/2021/017. The authors also acknowledge financial support from Spanish MICIN through Grant No. PID2019$-$109539GB$-$C43,  the Mar\'ia de Maeztu Program for Units of Excellence in R$\&$D (Grant No. CEX2018$-$000805$-$M), the Comunidad Aut\'onoma de Madrid through the Nanomag COST-CM Program (Grant No. S2018/NMT$-$4321). The theoretical modeling was performed on the high-performance computing facilities of the University of South Africa and the University of Alicante. Netherlands Organization for Scientific Research (NWO/OCW) supported the experiments. The CMD and DFT calculations in this paper were performed on the high-performance computing facilities of the University of South Africa and the University of Alicante.

%\section{References}
\bibliography{Jumptoedge.bib}

\newpage

\end{document}